\documentclass[aps,prl,twocolumn,superscriptaddress]{revtex4}

\usepackage{graphicx}
\usepackage{amsmath}
\usepackage{amssymb}

\begin{document}

\title{Orbital Feshbach Resonance in Alkali-Earth Atoms}
\author{Ren Zhang}
\affiliation{Institute for Advanced Study, Tsinghua University, Beijing, 100084, China}
\author{Yanting Cheng}
\affiliation{Institute for Advanced Study, Tsinghua University, Beijing, 100084, China}
\author{Hui Zhai}
\email{hzhai@tsinghua.edu.cn}
\affiliation{Institute for Advanced Study, Tsinghua University, Beijing, 100084, China}
\author{Peng Zhang}
\email{pengzhang@ruc.edu.cn}
\affiliation{Department of Physics, Renmin University of China, Beijing, 100872,
China}
\affiliation{Beijing Key Laboratory of Opto-electronic Functional Materials \&
Micro-nano Devices, 100872 (Renmin Univeristy of China)}

\date{\today }

\begin{abstract}
For a mixture of alkali-earth atomic gas in the long-lived excited state ${}^3P_0$ and the ground state ${}^1S_0$, in addition to nuclear spin, another ``orbital" index is introduced to distinguish these two internal states.  In this letter we propose a mechanism to induce Feshbach resonance between two atoms with different orbital and nuclear spin quantum numbers. Two essential ingredients are inter-orbital spin-exchanging process and orbital dependence of the Land\'e g-factors. Here the orbital degrees of freedom plays similar role as electron spin degree of freedom in magnetic Feshbach resonance in alkali-metal atoms. This resonance is particularly accessible for ${}^{173}$Yb system. The BCS-BEC crossover in this system requires two fermion pairing order parameters, and displays significant difference comparing to that in alkali-metal system.

\end{abstract}

\maketitle

Magnetic Feshbach resonance (MFR) is a powerful tool to tune interaction between atoms to strongly interacting regime, and plays a crucial role in cold atom physics \cite{FR}. For instance, for alkali atoms, each atom has an electronic spin $S=1/2$. The interaction between two atoms have different potentials depending on whether the total electronic spin is singlet or triplet. Thus, one can utilize the Zeeman energy to control their relative energy and to reach a scattering resonance. However, alkali-earth atom (like Sr) or alkali-earth-like atom (like Yb) has fully occupied outer shell, and their total electron spin is zero. Thus, it is conventional wisdom that there is no MFR in alkali-earth atoms at ground state. Instead, one can tune interaction by optical Feshbach resonance \cite{OFR}, but such a scheme suffers from strong atomic loss and heating. 

Another significant feature of alkali-earth atom is the existence of a long-lived excited state ${}^3P_0$, in which one electron is excited to $p$-orbital and the total electronic spin $S=1$. The dipole transition to ground state ${}^1S_0$ is ``spin-forbidden" and therefore the lifetime of excited state can be as long as a few seconds. This is used for atomic clock transition. Considering an atomic gas mixture of ${}^3P_0$ (denoted by $|e\rangle$) and ${}^1S_0$ (denoted by $|g\rangle$) states, in addition to nuclear spin degree of freedom, one introduces another so-called ``orbital" degree of freedom to label the internal state of atoms \cite{Jun,LENS,Bloch}.  Previously attentions have been paid to MFR between ${}^3P_2$ state and ${}^1S_0$ state due to anisotropic interactions, but these resonances are generally quite narrow \cite{P2}. Moreover, since $J=0$ for ${}^3P_0$, even such a MFR does not exist between ${}^3P_0$ and ${}^1S_0$. 

In this letter we propose an alternative mechanism that can lead to a Feshbach resonance (FR) between atoms in $|e\rangle$ and $|g\rangle$. Though this FR is also controlled by magnetic field, the mechanism of how it works is quite different from MFR in alkali atoms. Two essential ingredients are inter-orbital (nuclear-)spin-exchanging interactions, which has been observed in recent experiments \cite{Jun,LENS,Bloch}, and the small difference in the nuclear Land\'e g-factor $\delta g$ between different orbital states ($|e\rangle$ and $|g\rangle$) \cite{boyd}. The orbital degree of freedom plays the role as electronic spin in FR of alkali atoms, and we therefore name it as ``Orbital Feshbach resonance" (OFR).  

\begin{figure}[tp]
\includegraphics[width=3.4 in]
{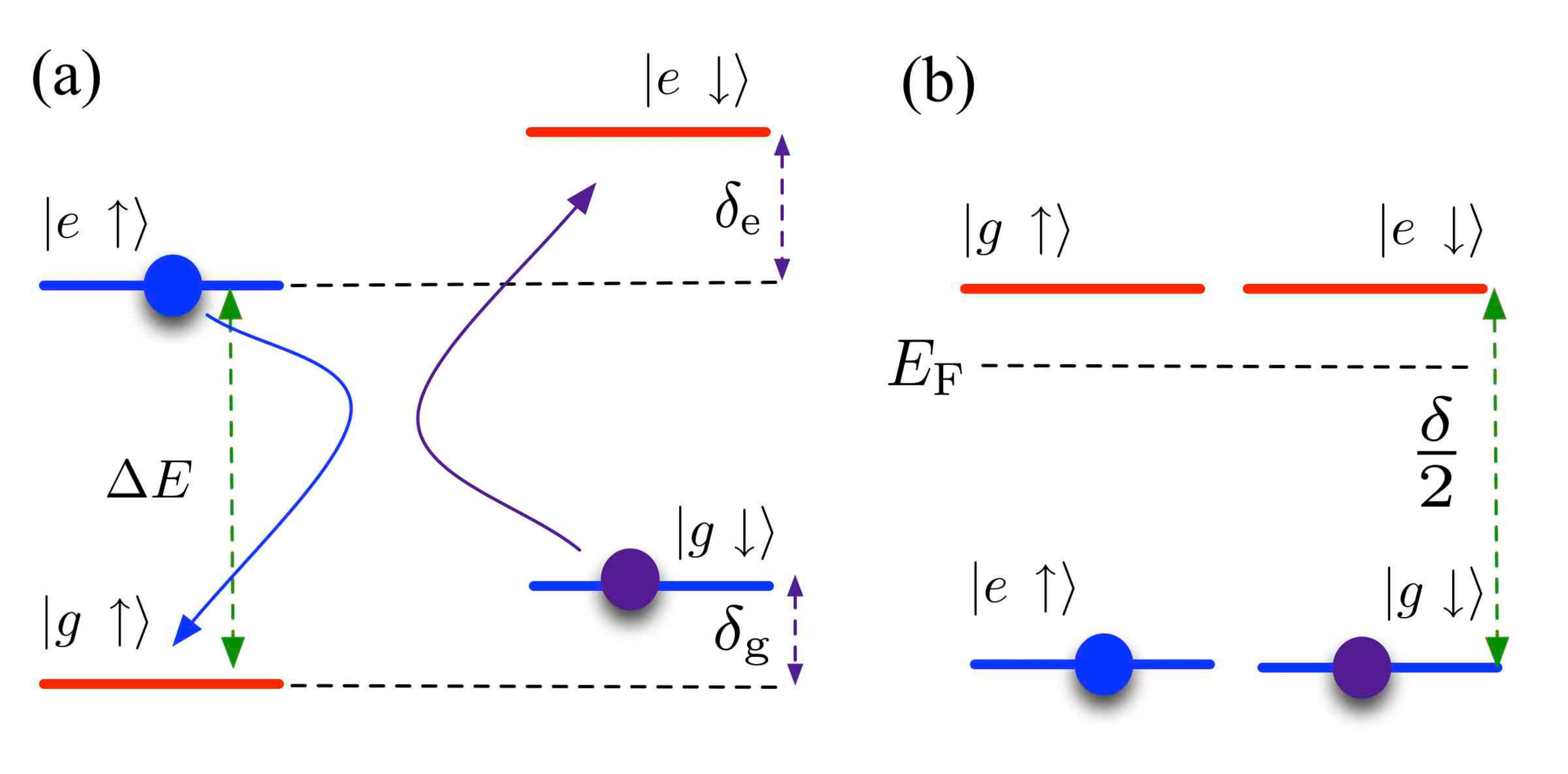}
\caption{(a) Original energy level diagram. $\Delta E$ denotes the excitation energy between $|e\rangle$ and $|g\rangle$. $\delta_\text{g}=Bg_\text{g}\mu_\text{B}$ and $\delta_\text{e}=Bg_\text{e}\mu_\text{B}$ are Zeeman energies of $|g\rangle$ and $|e\rangle$ states, respectively. Two states in open channel are occupied. Arrows indicate an inter-orbital (nuclear-) spin exchanging scattering process, which couples open channel $|g\downarrow;e\uparrow\rangle$ and closed channel $|g\uparrow;e\downarrow\rangle$. $\delta=\delta_\text{e}-\delta_\text{g}=B(\delta g)\mu_\text{B}$  is the Zeeman energy difference between two channels and $\delta g=g_\text{e}-g_\text{g}$ is the difference in Land\'e g-factor. (b) Reorganized energy level diagram for many-body Hamiltonian, in which open channel states appear in lower energy.  }
\label{level}
\end{figure}

\textit{Two-body Problem with Pseudo-Potential Approach.} For simplicity, we first illustrate the basic idea with pesudo-potential. And without loss of generality, we take two nuclear spin states ($m_I$ and $m_I+1$) denoted by $|\uparrow\rangle$ and $|\downarrow\rangle$. Their energy level diagram is shown in Fig. \ref{level}(a) and explained in the figure caption. We consider two atoms, either one in $|g\downarrow\rangle$ and the other in $|e\uparrow\rangle$ (denoted by open channel $|o\rangle=(|g\downarrow;e\uparrow\rangle-|e\uparrow;g\downarrow\rangle)/\sqrt{2}$), or either one in $|g,\uparrow\rangle$ and the other in $|e\downarrow\rangle$ (denoted by closed channel $|c\rangle=(|g\uparrow;e\downarrow\rangle-|e\downarrow;g\uparrow\rangle)/\sqrt{2}$). The threshold energy of these two channels differs by $\delta$ due to the difference in the Land\'e g-factor. For the relative motion between these two-atoms (with mass $m$), the non-interacting Hamiltonian is written as
\begin{equation}
\hat{H}_0=\left(-\frac{\hbar^2\nabla^2}{m}+\delta\right)|c\rangle\langle c|-\frac{\hbar^2\nabla^2}{m}|o\rangle\langle o|.
\end{equation}
The interaction part depends on whether orbital degree of freedom forms a singlet or triplet. For $s$-wave scattering we introduce two anti-symmetrized bases 
\begin{equation}
|\pm \rangle=\frac{1}{2}(|ge\rangle\pm |eg\rangle)(|\uparrow\downarrow\rangle\mp|\downarrow\uparrow\rangle)=\frac{1}{\sqrt{2}}(|c\rangle\mp |o\rangle), 
\end{equation}
the Huang-Yang pseudo-potential is diagonal in this bases with two different scattering lengths $a^+_\text{s}$ and $a^-_\text{s}$, 
\begin{equation}
\hat{V}=\left(\frac{4\pi\hbar^2}{m}\sum_{i=\pm}a_{\text{s}}^{(i)}|i\rangle\langle i|\right)\delta({\bf r})\frac{\partial}{\partial r}(r\cdot).\label{v}
\end{equation} 

When rotated into the $|o\rangle$ and $|c\rangle$ bases, the interaction potential $\hat{V}$ becomes
\begin{equation}
\hat{V}=\hat{V}_{0}\left(|o\rangle\langle o|+|c\rangle\langle c|\right)+\hat{V}_{1}\left(|c\rangle\langle o|+|o\rangle\langle c|\right),\label{v2}
\end{equation} 
where $V_j=\frac{4\pi\hbar^2}{m} a_{\text{s}j}\delta({\bf r})\frac{\partial}{\partial r}(r\cdot)$, and $a_{\text{s}0}$ denotes $(a^+_\text{s}+a^-_\text{s})/2$, and $a_{\text{s}1}$ denotes $(a^-_\text{s}-a^+_\text{s})/2$. The $\hat{V}_1$ term describes an inter-orbital spin exchanging process, as illustrated in Fig. \ref{level}, which couples the open and closed channels. A positive $a_{\text{s}0}$ is always associated with a bound state with binding energy $\varepsilon_{b}=-\hbar^2/(ma_{\text{s}0}^{2})$. Therefore, when $\delta\sim \varepsilon_{b}$, one will expect a scattering resonance in the open channel.   

The two-body wave function can be written as
\begin{equation}
\psi=\left[e^{i{\bf k}\cdot{\bf r}}+f_\text{o}(k)\frac{e^{ikr}}{r}\right]|o\rangle+f_\text{c}(k)\frac{e^{-\sqrt{m\delta/\hbar^2-k^{2}}r}}{r}|c\rangle. \label{psir}
\end{equation}
Solving the Schr\"odinger equation $(\hat{H}_0+\hat{V})\psi=E\psi$ with $E=\hbar^2 k^2/m$, one can find 
\begin{align}
&(1+ika_{\text{s}0})f_\text{o}(k)-a_{\text{s}1}\sqrt{\frac{m\delta}{\hbar^2}-k^{2}}f_\text{c}(k)+a_{\text{s}0}=  0;\label{e1}\\
&ika_{\text{s}1}f_\text{o}(k)+\left(1-a_{\text{s}0}\sqrt{\frac{m\delta}{\hbar^2}-k^{2}}\right)f_\text{c}(k)+a_{\text{s}1} = 0.\label{e2}
\end{align}
Straightforward calculation yields the scattering length $a_\text{s}$ in the open channel as 
\begin{equation}
a_{\text{s}}=-f_\text{o}(k=0)=\frac{-a_{\text{s}0}+\sqrt{m\delta/\hbar^2}(a^2_{\text{s}0}-a^2_{\text{s}1})}{a_{\text{s}0}\sqrt{m\delta/\hbar^2}-1}.\label{aa}
\end{equation} 

\textit{Two-body Problem with Finite Range Potential.} We can further more rigorously demonstrate the OFR with a coupled two-channel model with finite range $r_0$. When $r>r_0$, two atoms are non-interacting, and the zero-energy $s$-wave wave function $\psi=u(r)/r$ with
\begin{equation}
u(r)=\alpha \exp\left(-\sqrt{\frac{m\delta}{\hbar^2}}r\right)|c\rangle+\beta (r-a_\text{s})|o\rangle.  \label{out}
\end{equation}
For $r<r_0$, the Hamiltonian is written as
\begin{equation}
\hat{H}=\sum\limits_{i=+,-}\left(-\frac{\hbar^2\nabla^2}{m}+V^i({\bf r})\right)|i\rangle\langle i|,
\end{equation}
where we have assumed that $\delta$ is much smaller than energy scale of short-range potential $V^i({\bf r})$ such that it can be safely ignored in this regime. Each $V^i({\bf r})$ ($i=+,-$) corresponds to  an $s$-wave scattering length $a^i_\text{s}$, that is to say, the wave function $\psi^i=u^i(r)/r$ in $r<r_0$ regime satisfies the boundary condition 
$u^{i\prime}(r)/u^{i}(r)|_{r=r_{0}}=1/(r_0-a_{s}^{i})$.
In $r<r_0$ regime the wave function can be written in a general form 
\begin{align}
&u(r)=u^+(r)|+\rangle+A u^-(r)|-\rangle \nonumber\\
&=\frac{u^{+}(r)+Au^{-}(r)}{\sqrt{2}}|c\rangle+\frac{-u^{+}(r)+Au^{-}(r)}{\sqrt{2}}|o\rangle \label{in}
\end{align}

\begin{figure}[tp]
\includegraphics[width=3.0 in]
{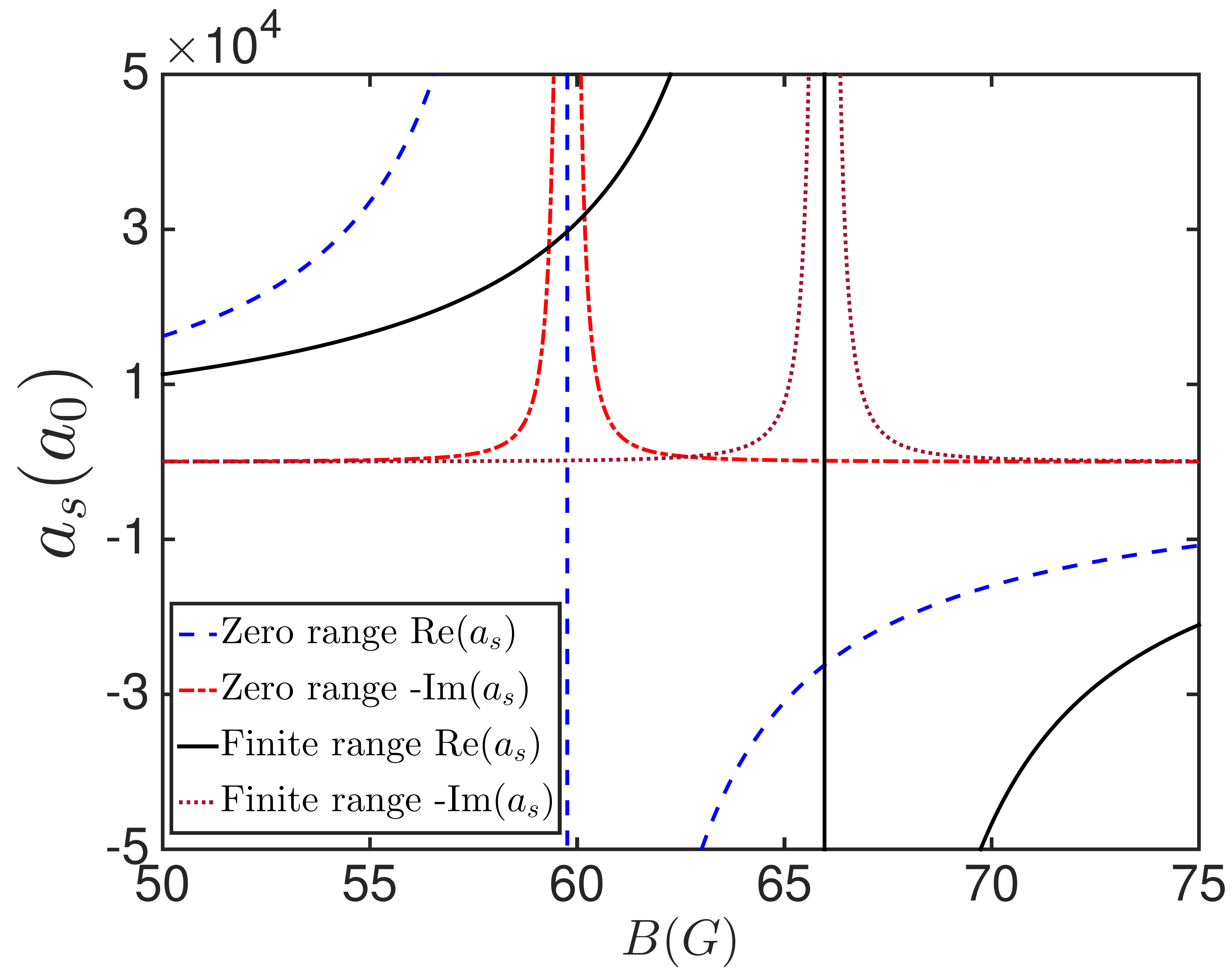}
\caption{Scattering length $a_\text{s}$ between $|e\uparrow\rangle$ and $|g\downarrow\rangle$ as a function of magnetic field for ${}^{173}$Yb atom. Blue dashed line and red dash-dotted line is the real and imaginary part of $a_\text{s}$ obtained from the zero-range pseudo-potential. Solid line and red dotted line are $a_\text{s}$ obtained from the finite range potential. Here we take $(\delta g)
\mu_\text{B}=2\pi\hbar\times112{\rm Hz}/{\rm G}$,
$a^{+}_\text{s}=3300a_{0}-i0.78a_{0}$, $a^{-}_\text{s}=219.5a_{0}$
\cite{Bloch,LENS,location} and $r_0$ is taken as van der Waal length, which equals to $84.8a_{0}$ \cite{martin}, with $a_{0}$ the Bohr's radius.   }
\label{scatteringlength}
\end{figure}

By matching boundary conditions between wave functions Eq. \ref{out} and Eq. \ref{in} at $r=r_0$ for $|o\rangle$ and $|c\rangle$ channels independently, and 
utilizing the boundary condition for each $u^i$, one can obtain $A$ and $a_\text{s}$, where
\begin{equation}
a_\text{s}=\frac{-a_{\text{s}0}+\sqrt{m\delta/\hbar^2}\left[(a_{\text{s}0}^{2}-a_{\text{s}1}^{2})-r_{0}a_{\text{s}0}\right]}{\sqrt{m\delta/\hbar^2}\left(a_{\text{s}0}-r_{0}\right)-1}. \label{asr0}
\end{equation}
In the limit $r_0\rightarrow 0$, Eq. \ref{asr0} recovers the result of Eq. \ref{aa}. Thus we have demonstrated the OFR phenomena. Eq. \ref{aa} and Eq. \ref{asr0} also show that the difference between $a^-_\text{s}$ and $a^+_\text{s}$, i.e. $a_{\text{s}1}\neq 0$, is crucial, as $a_\text{s}$ becomes a constant as $a_{\text{s}0}$ if one sets $a_{\text{s}1}=0$ in Eq. \ref{aa} and Eq. \ref{asr0}. 

\begin{figure}[tp]
\includegraphics[width=2.4 in]
{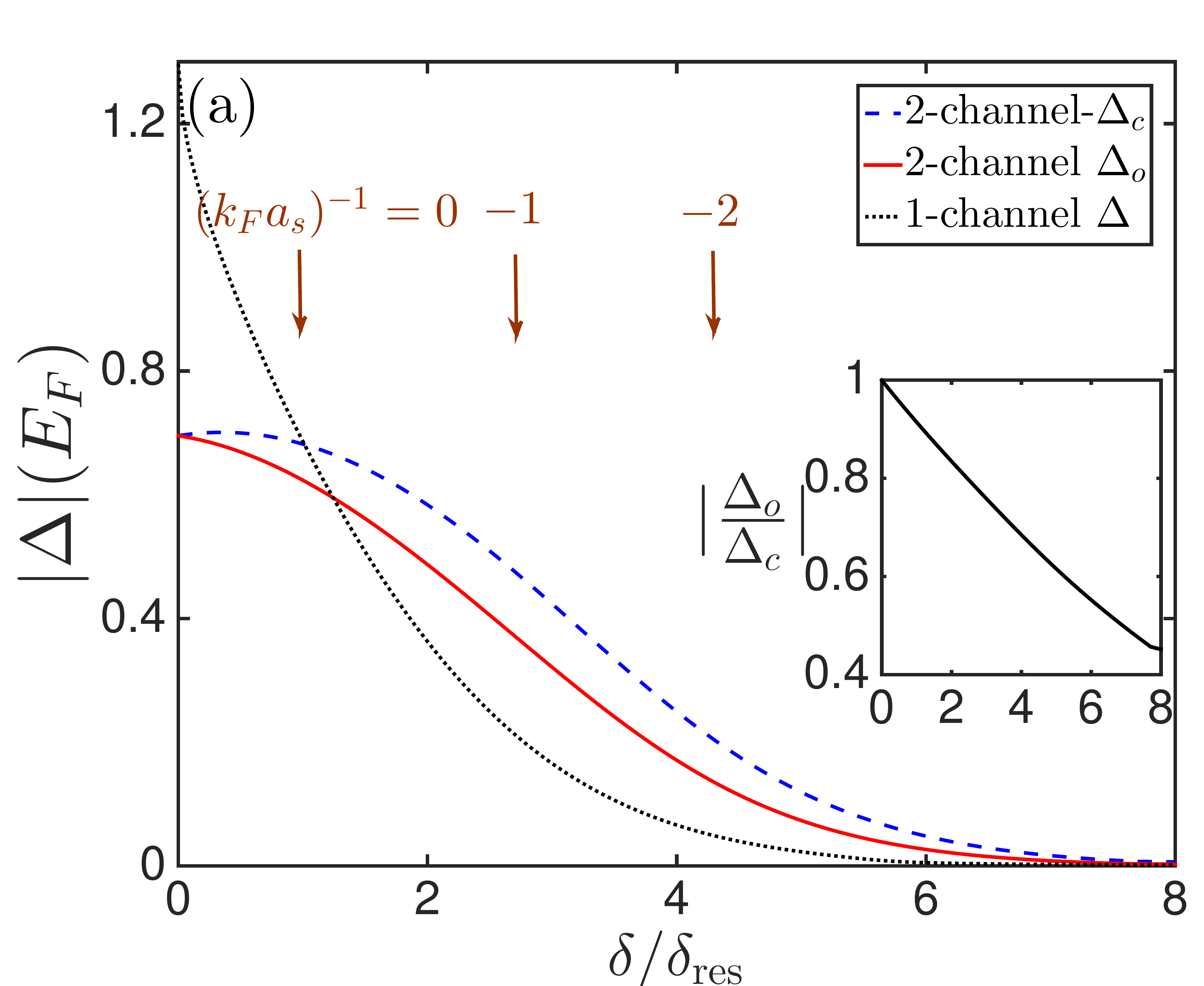}
\includegraphics[width=2.4 in]
{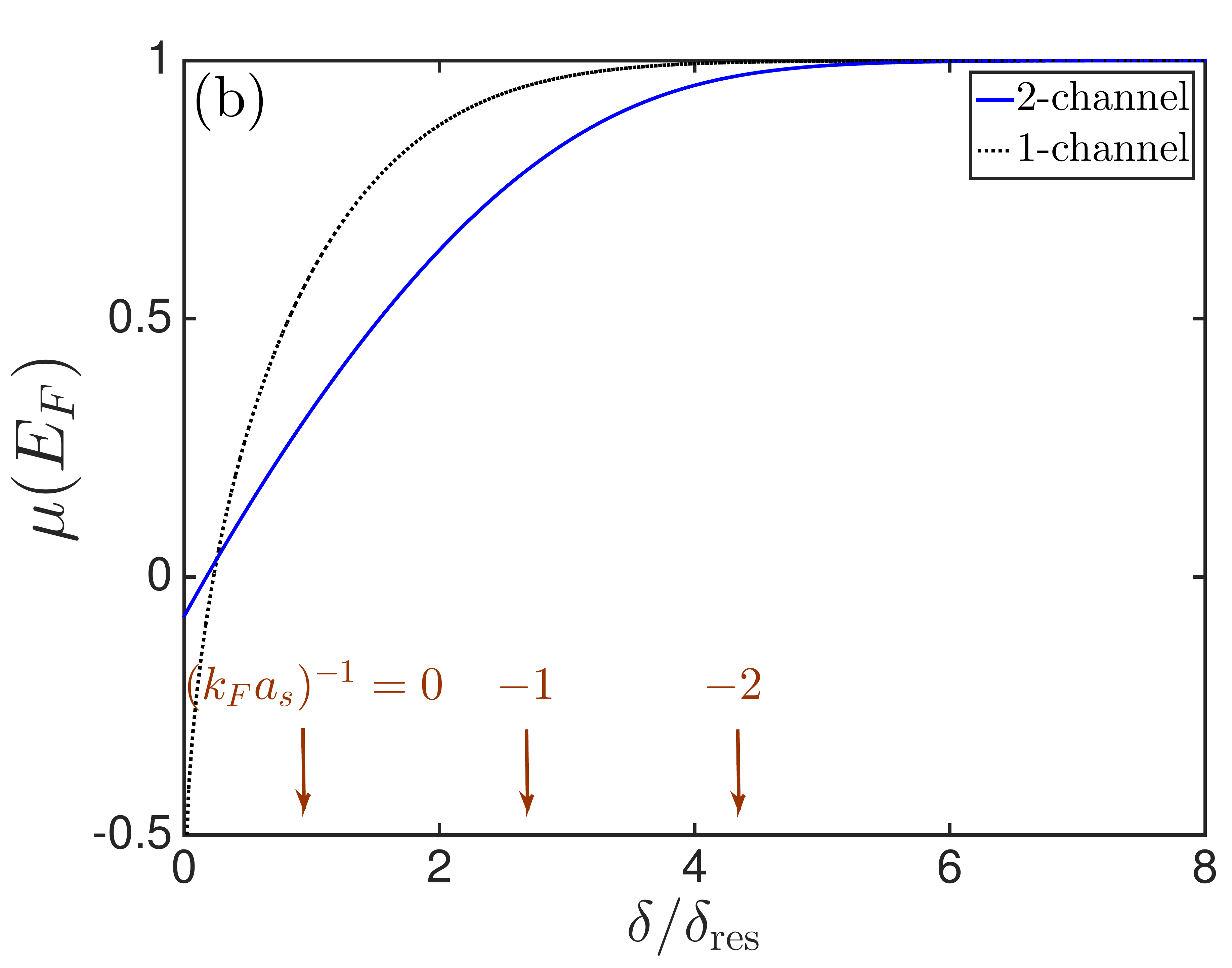}
\caption{Pairing order parameters (a) and chemical potential (b) as a function of $\delta/\delta_\text{res}$ across an OFR, with $\delta_\text{res}$ given by Eq. \ref{hres} ($r_0=0$). In (a), red solid and blue dashed lines are pairing order parameters in the open and closed channel, respectively. For different $\delta/\delta_\text{res}$, corresponding values of $1/(k_\text{F}a_\text{s})$ are marked by arrows, where $a_\text{s}$ is the open channel scattering length. Here $a^+_\text{s}$ and $a^-_\text{s}$ are taken from ${}^{173}$Yb atom, as described in caption of Fig. \ref{scatteringlength}, and $n=k^3_\text{F}/(3\pi^2)$ is taken as $5\times 10^{13}\text{cm}^{-3}$ in our calculation. For comparison, black dotted lines in (a) and (b) show pairing order parameter and chemical potential for single-channel BCS-BEC crossover with same $1/(k_\text{F}a_\text{s})$. }
\label{mean-field}
\end{figure}

\textit{Orbital Feshbach Resonance.} Eq. \ref{asr0} shows that $a_\text{s}$ diverges when 
\begin{equation}
\delta=\delta_{\rm res}=\frac{\hbar^2}{m(a_{\text{s}0}-r_{0})^{2}}. \label{hres}
\end{equation}
This determines the position of OFR. It also indicates that the precise location of OFR may be altered by short-range details. In addition, we also obtain that $a_\text{s}$ has a zero-crossing when 
\begin{equation}
\delta=\delta_{0}=\frac{\hbar^2}{m(a_{\text{s}0}-r_{0}-a_{\text{s}1}^{2}/a_{\text{s}0})^{2}}.
\end{equation}
Since $\delta=B(\delta g)\mu_\text{B}$, we take $a^+_\text{s}$, $a^-_\text{s}$, $r_0$ and $(\delta g)\mu_\text{B}$ from ${}^{173}$Yb atom measurement reported so far, and plot $a_\text{s}$ as a function of magnetic field $B$, as shown in Fig. \ref{scatteringlength}. ${}^{173}$Yb is a unique system to observe an OFR, since $a_\text{s}^{+}$ is as large as $\sim 10^3a_{0}$. Therefore, it does not require large magnetic field to reach the OFR. Otherwise, if the scattering length is small, it requires a much high magnetic field ($10^3-10^4$G for ${}^{87}$Sr) to reach this resonance.  On the other hand, since the essential ingredient of inter-orbit spin-exchanging process also exists between ${}^3P_1$ (or ${}^3P_2$) state and ${}^1S_0$, OFR can also exist in these mixtures, where the $g$-factor difference is quite large, and it does not rely  on a large scattering length at zero-field. Nevertheless, the scattering behavior between these states is also more complicated, and the details are left for future studies. 

Regarding loss nearby OFR, there are two mechanisms. First, the inelastic two-body loss, described by the imaginary part of scattering length, is also shown in Fig. \ref{scatteringlength}.  Secondly, the shallow bound state is also subjected to three-body loss, which cannot be rigorously captured by our two-body calculation. Nevertheless, we phenomenologically include this effect by introducing a decay term on the shallow bound state. We find that it will not affect the real part of scattering length, as long as the bound state has a reasonable lifetime, but only enhance its imaginary part \cite{supple}.

Here we would like to contrast the OFR in alkali-earth atom with MFR in alkali-metal atom. In MFR, when two atoms interact at short distance, the interaction potential are different for total electronic {\it spin} singlet and triplet. While for OFR, the interaction potentials are distinguished by {\it orbital} singlet or triplet. In MFR, the coupling between two channels is due to hyperfine interaction. While in OFR, orbital dependent Land\'e g-factor can be viewed as coupling between orbital and nuclear spin, because of which two channels are coupled. In this analogy, orbital degree of freedom in OFR plays the same role as electronic spin degree of freedom in MFR. 

\textit{Two Order-Parameters BCS-BEC Crossover.} Since $(\delta g)\mu_\text{B}$ is five orders of magnitude smaller than the $g_e\mu_\text{B}$, comparing OFR with a MFR in the same magnetic field regime, the energy separation between open and closed channel is much larger in the MFR case than that in the OFR case. In a MFR case, this energy separation is a few orders of magnitude larger than the Fermi energy. Therefore, in a BCS-BEC crossover theory studied before \cite{Stringari}, one can either start with a single-channel model only, or with a two-channel model but only including the bound state of the closed channel. The scattering states in the closed channel are never important.  However, the situation in the OFR case is considerably different. Considering a typical density of Fermi gas, $\delta$ is comparable or can be even smaller than the Fermi energy. Thus, we have to take into account scattering states in both open and closed channels. This requires introducing two self-consistent paring order parameters for open and closed channels, respectively. Below we shall present such a formalism for crossover across an OFR. 

We consider the situation that in the non-interacting limit, total $N$ fermions are equally populated in the two states in the open channel ($|e\uparrow\rangle$ and $|g\downarrow\rangle$). We note that both $N_{e}=N_{e\uparrow}+N_{e\downarrow}$ and $N_\uparrow=N_{g\uparrow}+N_{e\uparrow}$ are good quantum numbers. Subtracting the Hamiltonian with a constant term $(\Delta E+\delta/2)N_e-(\delta_e+\delta_g)N_\uparrow/2$, one can show the level diagram can be reorganized as shown in Fig. \ref{level}(b), such that the two states in the open channel appear in the lower energy, with an energy separation of $\delta/2$ below the two states in the closed channel. Then in the non-interacting limit (say, $\delta=\delta_0$), $\delta/2$ is larger than the Fermi energy so that only open channel is equally populated. The many-body Hamiltonian can be written as 
\begin{align}
&\hat{H}=\hat{H}_{0\text{o}}+\hat{H}_{0\text{c}}+\frac{g_+}{2}\hat{A}^\dag_{+}\hat{A}_++\frac{g_-}{2}\hat{A}^\dag_{-}\hat{A}_-,\\
&\hat{H}_{0\text{o}}=\sum_{{\bf k}}\varepsilon_{\bf k}(c_{g\downarrow {\bf k}}^{\dagger}c_{g\downarrow {\bf k}}+c_{e\uparrow {\bf k}}^{\dagger}c_{e\uparrow {\bf k}})\\
&\hat{H}_{0\text{c}}=\sum_{{\bf k}}\left(\varepsilon_{\bf k}+\frac{\delta}{2}\right)(c_{g\uparrow {\bf k}}^{\dagger}c_{g\uparrow {\bf k}}+c_{e\downarrow {\bf k}}^{\dagger}c_{e\downarrow {\bf k}}),
\end{align}
where $\varepsilon_{\bf k}=\hbar^2{\bf k}^2/(2m)-\mu$, and  
\begin{align}
&\hat{A}_{+}=\sum_{{\bf k}}(c_{g\uparrow\bf -k}c_{e\downarrow\bf k}-c_{g\downarrow\bf -k}c_{e\uparrow\bf k})\\
&\hat{A}_{-}=\sum_{{\bf k}}(c_{g\uparrow\bf -k}c_{e\downarrow\bf k}+c_{g\downarrow\bf -k}c_{e\uparrow\bf k}).
\end{align}

Now we defined two order parameters as $\Delta_{+}=g_{+}\langle \hat{A}\rangle/2$ and $\Delta_{-}=g_{-}\langle \hat{A}\rangle/2$, and we can perform mean-field decoupling of the interaction term which leads to 
\begin{align}
\hat{H}_\text{MF}=&\hat{H}_{0\text{o}}+\hat{H}_{0\text{c}}+(\Delta_{+}\hat{A}_{+}+\text{h.c.})\nonumber\\
&+(\Delta_{-}\hat{A}_{-}+\text{h.c.})-\frac{2|\Delta_{+}|^{2}}{g_{+}}-\frac{2|\Delta_{-}|^{2}}{g_{-}}.
\end{align}
Following the standard BCS theory to diagnoalized $\hat{H}_\text{MF}$ with the Bogoliubov transformation and minimizing the ground state energy with respect to both $\Delta_+$ and $\Delta_{-}$ \cite{Book}, we reach two coupled gap equations 
\begin{align}
\left[\frac{\frac{\Delta_\text{o}}{\Delta_\text{c}}-1}{\frac{4\pi\hbar^{2}a_{s}^{+}}{m}}-\frac{1+\frac{\Delta_\text{o}}{\Delta_\text{c}}}{\frac{4\pi\hbar^{2}a_{s}^{-}}{m}}\right]&=\sum_{\bf k}\frac{1}{\sqrt{\left(\varepsilon_{{\bf k}}+\frac{\delta}{2}\right)^{2}+|\Delta_\text{c}|^{2}}}-\frac{2m}{\hbar^2 {\bf k}^2}\nonumber\\
\left[\frac{\frac{\Delta_\text{c}}{\Delta_\text{o}}-1}{\frac{4\pi\hbar^{2}a_{s}^{+}}{m}}-\frac{1+\frac{\Delta_\text{c}}{\Delta_\text{o}}}{\frac{4\pi\hbar^{2}a_{s}^{-}}{m}}\right]&=\sum_{\bf k}\frac{1}{\sqrt{\varepsilon_{{\bf k}}^{2}+|\Delta_\text{o}|^{2}}}-\frac{2m}{\hbar^2{\bf k}^2} \label{gap}
\end{align}
where $\Delta_\text{o}=\Delta_{-}-\Delta_{+}$ and $\Delta_{\text{c}}=\Delta_{-}+\Delta_{+}$ are pairing order parameters in the open and closed channels, respectively. Here, in contrast to usual BCS-BEC crossover where scattering length is the tunable control parameter, here both $a^+_\text{s}$ and $a^-_\text{s}$ are fixed. Instead, $\delta$ is the tunable parameter to control the crossover. Moreover, the number equation also includes contribution from both channels 
\begin{equation}
N=\sum_{\bf k}\left(2-\frac{\varepsilon_{{\bf k}}}{\sqrt{\varepsilon_{\bf k}^{2}+|\Delta_{\text{o}}|^{2}}}-\frac{\varepsilon_{{\bf k}}+\frac{\delta}{2}}{\sqrt{\left(\varepsilon_{{\bf k}}+\frac{\delta}{2}\right)^{2}+|\Delta_{\text{c}}|^{2}}}\right).\label{number}
\end{equation}

Solving gap equation Eq. \ref{gap} with Eq. \ref{number}, we find $\Delta_{\text{o}}$ has a $\pi$ phase difference from $\Delta_\text{c}$, and we determine both $|\Delta_{\text{o}}|$, $|\Delta_\text{c}|$ and $\mu$ as a function of $\delta$, as shown in Fig. \ref{mean-field}. We find (i) when $\delta \gg \delta_\text{res}$, the system is away from OFR. In this case we find small $|\Delta_{\text{o}}|/E_\text{F}$, $|\Delta_\text{c}|/E_\text{F}$ and $\mu\rightarrow E_\text{F}$, which is the typical behavior in the BCS regime. (ii) As $\delta\rightarrow \delta_\text{res}$, both $|\Delta_{\text{o}}|$ and $|\Delta_\text{c}|$ increase rapidly toward the same order as $E_\text{F}$ and meanwhile, $\mu$ decreases. This feature is qualitatively consistent with a crossover from BCS to the unitary regime. (iii) While when $\delta<\delta_\text{res}$ and $\delta\rightarrow 0$, both pairing gaps saturate instead of continuously increasing toward deep BEC limit. This is consistent with that $a_\text{s}$ finally saturates to $a_{\text{s}0}$.

We also plot the ratio $|\Delta_\text{o}/\Delta_\text{c}|$ in the inset of Fig. \ref{mean-field}(a), we find when $\delta \gg \delta_\text{res}$, this ratio decreases toward zero, and thus the L.H.S. of Eq. \ref{gap} diverges as it depends on $\Delta_\text{c}/\Delta_\text{o}$. Effectively, if one compares Eq. \ref{gap} with a single-channel BCS-BEC gap equation \cite{Book}, this is equivalent to that an open channel scattering length decreases toward zero. When $\delta\rightarrow \delta_\text{res}$, $|\Delta_\text{o}/\Delta_\text{c}|\rightarrow (a^+_\text{s}-a^-_\text{s})/(a^+_\text{s}+a^-_\text{s})$ ($\sim 0.875$ for the scattering lengths we use). The L. H. S. of Eq. \ref{gap} approaches zero, indicating an divergent effective scattering length. Finally when $\delta\rightarrow 0$, two channels become degenerate and thus this ratio approaches unity. We remark that this mean-field calculation does not use the results from two-body calculation above. It is an independent many-body calculation, while the results are qualitatively consistent with two-body results. 

On the other hand, for the typical density we consider, the quantitative behavior is quite different from the single channel BCS-BEC crossover. We perform a single-channel mean-field calculation, in which we only keep the two states in the open channel and use open channel scattering length $a_\text{s}(\delta)$ given by two-body result of Eq. \ref{aa}. The result is shown by the dotted line in Fig. \ref{mean-field} and compared with the two-gap theory presented above. Remarkably, we find that in the BCS regime, the pairing in the two-gap theory is stronger than that in the single channel model. While if we lower the density so that the Fermi energy becomes much lower than $\delta/2$, the results gradually converges to the single channel result \cite{Lianyi_He}.  

\textit{Outlook.} Our predication of OFR opens an avenue for studying strongly interacting physics in alkali-earth atomic gases. The two-gap Fermi superfluid is reminiscent of two-gap superconductor. Further studies including Gaussian fluctuations can reveal effect of finite life time of Cooper pairs, superfluid transition temperature and the Leggett mode (relative phase mode between two gaps) in strongly interacting regime, which is left for future investigations. Moreover, by coupling $|e\uparrow\rangle$ and $|g\downarrow\rangle$ states with a laser, one can create spin-orbit coupling between them, which avoids heating from spontaneous emission as in Raman scheme. Our OFR increases attractive interaction between them and can help to reach a topological superfluid phase in this system.  

\textit{Acknowledgement.} We would like to thank Jun Ye, Leonardo Fallani, Yoshiro Takahashi, Simon F\"olling, and Immanuel Bloch for discussion atomic properties of alkali-earth atoms. This work is supported by Tsinghua University Initiative Scientific Research Program, NSFC Grant No. 11174176(HZ), No. 11325418 (HZ), No. 11222430 (P.Z.), No. 11434011 (PZ) and NKBRSFC
under Grant No. 2011CB921500 (HZ) and No. 2012CB922104 (P.Z.).


\begin{thebibliography}{99}
\bibitem{location} The exact location of OFR will be determined by the accurate value of $a^+_\text{s}$ and $a^-_\text{s}$.
\end{thebibliography}
\end{document}